\documentclass[12 pt]{article}
\usepackage{times,bm}
\usepackage[affil-it,]{authblk}
\usepackage[pagebackref=false,colorlinks=true,linkcolor=blue,citecolor=blue,urlcolor=blue]{hyperref}
\usepackage{geometry}
\usepackage{graphicx}
\usepackage{cite}
\usepackage{amsmath}
\usepackage{amsfonts}
\usepackage{amssymb}
\usepackage{color}
\usepackage{float}
\usepackage{multirow,multicol}
\usepackage{tabulary}
\usepackage[flushleft]{threeparttable}
\usepackage{pgfplots,subfigure}
\usepackage{xcolor}
\numberwithin{equation}{section}
\usepackage{tikz}
\usepackage{ulem}
\usepackage{hyperref}
\usepackage{nameref}
\usepackage{comment}

\usepgflibrary{arrows}
\usetikzlibrary{shapes.callouts}
\tikzset{
	level/.style   = { thick, },
	connect/.style = { dotted, red   },
	notice/.style  = { draw, rectangle callout, callout relative pointer={#1} },
	label/.style   = { text width=2cm }
}

\usepackage{letltxmacro}
\makeatletter
\let\oldr@@t\r@@t
\def\r@@t#1#2{%
	\setbox0=\hbox{$\oldr@@t#1{#2\,}$}\dimen0=\ht0
	\advance\dimen0-0.2\ht0
	\setbox2=\hbox{\vrule height\ht0 depth -\dimen0}%
	{\box0\lower0.4pt\box2}}
\LetLtxMacro{\oldsqrt}{\sqrt}
\renewcommand*{\sqrt}[2][\ ]{\oldsqrt[#1]{#2}}
\makeatother
\begin{document}
	
\baselineskip=18 pt
\begin{center}
{\Large {\bf{ Interaction of the generalized Duffin–Kemmer–Petiau equation with a non-minimal coupling under the cosmic rainbow gravity } }}		
\end{center}
\vspace{.5cm}
	
\begin{center}
		
{ M. Hosseinpour $^{1}$}\,\,,\,\,{ H. Hassanabadi $^{1,2}$} \,\,,\,\,{  J. Kříž $^{2}$}\,\,,\,\,{ S. Hassanabadi $^{2}$}\,\,, and \,\,{B.C. L\"utf\"uo\u glu $^{2,3,}$\footnote{Email:  bclutfuoglu@akdeniz.edu.tr (Corresponding author) }}
\\
{\it $^{1}$ Faculty of Physics, Shahrood University of Technology, Shahrood, Iran \\ P. O. Box : 3619995161-316.}\\
\small \textit {\it $^{2}$ Department of Physics, University of Hradec Kr$\acute{a}$lov$\acute{e}$,
Rokitansk$\acute{e}$ho 62, 500 03 Hradec Kr$\acute{a}$lov$\acute{e}$, Czechia.\\
\it $^{3}$ Department of Physics, Akdeniz University, Antalya, 07058 Turkey. 	}
\end{center}
	
\vspace{.5cm}
\begin{abstract}
{In this study, we survey the   generalized Duffin–Kemmer–Petiau oscillator containing a non-minimal coupling interaction in the context of rainbow gravity in the presence of the cosmic topological defects in space-time. In this regard, we intend to investigate relativistic quantum dynamics  of a spin-0 particle under the modification of the dispersion relation according to the Katanaev-Volovich geometric approach. Thus, based on the geometric model, we study the aforementioned bosonic system under the modified background by a few rainbow functions. In this way, by using an analytical method, we acquire energy eigenvalues and corresponding wave functions corresponding to each scenario.  Regardless of rainbow gravity function selection, the energy eigenvalue can present symmetric, anti-symmetric, and symmetry breaking characteristics. Besides, one can see that the deficit angular parameter plays an important role in the solutions.}	\end{abstract}
	
	
\section{Introduction}
In physics, quantum gravity is a challenging problem that has garnered attention for many years. One of the semi-classical approaches to this phenomenon is given with the rainbow gravity(RG) model. There, another invariant apart from the velocity of light, namely the Planck energy,  is taken into account as an invariant energy scale set \cite{1,2,3,4}. In this approach, quantum corrections are assumed to depend on space-time via the metric tensor,  and vary with the energy of the probing particles \cite{5,6 ,7 ,8, 9}. More precisely, in this context, it is shown that the nonlinear representation of the Lorentz transformations in momentum space gives an energy-dependent space-time in the relativistic regime, so that, the relativistic dispersion relation changes according to the investigated RG models \cite{10}. Therefore, particles with different energies and wavelengths are not affected by space-time structure in the same way \cite{11}.

Recently, the semi-classical approach of the RG is employed to examine the thermodynamic quantities of  black holes \cite{12,13,14,15,16} and the structure of neutron stars \cite{17}. In particular, it is used in the space-times described by the  Schwarzschild \cite{18,19,20} and Friedmann-Robertson-Walker metrics
\cite{21,22,23,24}. Furthermore,  the RG is also  handled to describe the particle dynamics by considering the modified dispersion relation in the relativistic quantum mechanics \cite{25} and field theories \cite{26}. Besides, in the context of rainbow gravity, the relativistic behavior of spin-zero and spin-one bosons have been investigated in a topologically trivial G\"{o}del-type space-time and cosmic string space-time, respectively in \cite{KangalPS2021,SogutAP2021}.

In this manuscript, we intend to examine a Duffin–Kemmer–Petiau (DKP) oscillator \cite{Kemmer1938,Duffin1938,Kemmer1939,Petiau1936, Moshinsky1989,ZareEJMPA2020} in a cosmic string space-time under the presence of high-order correction terms which arises from the modified dispersion relation related to the  RG approach.  Here, the DKP oscillator is defined  through a non-minimal coupling of the linear potential energy term. In this way, the DKP equation mimics a harmonic oscillator equation in the weak-coupling limit \cite{27,28,29,30,31,32,33,34,35}. As is well known, one can obtain solutions to spin-0 and spin-1 particles separately out of a DKP oscillator problem. For example, Guo et al. in \cite{30},  and later, Yang et al. in \cite{31} explored solutions to the DKP oscillator with spin-0 particles in three-dimensions for commutative and noncommutative space, respectively. In \cite{32}, minimal length formalism is taken into account and solutions of the bosonic wave equation are given.  De Melo  et al. proposed a higher- dimensional formalism of  the Galilean covariance to the non-commutative DKP oscillator to obtain an analytic solution  \cite{36}. Hence, we observe that the interest in the DKP oscillator and its solutions is increasing day by day even in flat \cite{33,34,35,36,37,39,39a} and in curved space-times
\cite{40,41,42,43,44,45,46}. Considering all these studies, we are motivated to investigate the dynamics of the spin-0 bosonic vector field with a non-minimal coupling in the context of RG under the presence of the cosmic string space-time with topological defects.

We organize the manuscript as follows: At first, we present a brief review of the RG approach. Then, we introduce the DKP oscillator in a cosmic string background in the context of RG by defining the necessary tetrad basis and spin connections. Then, we introduce novelty to the manuscript by embedding the non-minimal coupling interaction into the generalized spin-0 DKP equation.  We derive the radial equations and their energy eigenvalue functions with the corresponding wave functions for three different pairs of RG functions with analytical methods. For each cases, we evaluate the solutions numerically and interpret them graphically. Finally, we conclude the article with a brief conclusion.
	
	%%%%% SEC 2

\section{The cosmic rainbow gravity}
	
We start by  describing space-time background through the metric of a cosmic string in the following form \cite{45,46}:
\begin{equation}\label{le1}
	ds^{2}=-dt^{2}+dr^{2}+\alpha^{2}r^{2}d\varphi^{2}+dz^{2},
\end{equation}	
where the coordinates can vary only in the following intervals: 	 $-\infty < z < \infty$, $r \geq 0$, and $0\leq \varphi \leq 2 \pi$. Hereafter, for simplicity we employ the natural units where $\hbar=1$ and $c=1$. Here, $\alpha$, so-called the angular parameter, links to the cosmic string's linear mass density, $\mu$, via  $\alpha=1-4G\mu$, with the Newton’s constant $G$. The angular parameter varies in the  interval $(0,1)$ and corresponds to a deficit angle by $\gamma=2 \pi (1 - \alpha)$.	
	
In the context of RG,  probe particles are assumed to influence the space-time background. Therefore, the metrics become energy-dependent so that a variety of metrics can be taken into account. At high energy scales, this assumption leads to a modification in the dispersion relation of the probe particles \cite{5}.
	\begin{equation}\label{MDR}
	E^{2}g_{0}^{2}\left(\frac{E}{E_{p}}\right)-p^{2}g_{1}^{2}\left(\frac{E}{E_{p}}\right)=M^{2}.
\end{equation}
Here, $E_{p}$ and $M$ correspond to the Planck energy and the mass of the probe particle, respectively.  By virtue of Eq. \eqref{MDR} the mass parameter tunes a mutual relation between the probe particles and the space-time background. The functions $g_{0}(x)$ and $g_{1}(x)$ are called rainbow functions, where $x=\frac{E}{E_{p}}$ is the ratio of the energy of the probe particle to the Planck energy. At the low-energy scale where $E\rightarrow 0$, thus $x\rightarrow 0 $ and
\begin{equation}\label{limRF}
g_{0}(0)=g_{1}(0)=1,
\end{equation}
so that the ordinary dispersion relation is recovered. In the framework of the rainbow gravity, the space-time expressed in Eq. \eqref{le1} is generalized to the following form \cite{5,11,12}
\begin{equation}\label{le2}
ds^{2}=-\frac{1}{g_{0}^{2}(x)}dt^{2}+\frac{1}{g_{1}^{2}(x)}\left[dr^{2}+\alpha^{2}r^{2}d\varphi^{2}+dz^{2}\right],
\end{equation}
in which the signature of the line element Eq. \eqref{le2} is $(-,+,+,+)$.
Thus, we retrieve the metric tensor,  $\mathrm{g}_{\mu\nu}$, out of the line element Eq. \eqref{le2} as
\begin{equation}\label{tmgd}
{\mathrm{g}_{\mu\nu}}=\mathrm{diag}\left(-\frac{1}{g_{0}(x)^{2}},\frac{1}{g_{1}(x)^{2}},\frac{\alpha^{2}r^{2}}{g_{1}(x)^{2}},\frac{1}{g_{1}(x)^{2}}\right),\qquad {\mu,\nu=t,r,\varphi,z}.
\end{equation}
%\begin{align}\label{tmgd}
%{\mathrm{g}_{\mu\nu}}=\left( \begin{matrix}
%-\frac{1}{g_{0}(x)^{2}}
%&0&0&0\\
%0&\frac{1}{g_{1}(x)^{2}}&0&0\\
%0&0&\frac{\alpha^{2}r^{2}}{g_{1}(x)^{2}}&0\\
%0&0&0&\frac{1}{g_{1}(x)^{2}}
%\end{matrix}\right),\qquad %{\mu,\nu=t,r,\varphi,z}.
%\end{align}
 In the rest of the manuscript, we examine the following three pairs of rainbow functions:
\begin{enumerate}
    \item The ones that are studied in Refs. \cite{25,26,Deng2018,Ashour2016}
    \begin{equation}\label{RFm1}
g_{0}(x)=g_{1}(x)=\frac{1}{1-\varepsilon x}.
\end{equation}
\item The ones that are used in Refs. \cite{6,21,22,24}
 \begin{equation}\label{RFm2}
g_{0}(x)=1,  \quad \quad g_{1}(x)=\sqrt{1-\varepsilon x^{2}}.
\end{equation}
\item The ones that are employed in Refs. \cite{6,21,22,24,25}
\begin{equation}\label{RFm3}
g_{0}(x)=\frac{e^{\varepsilon x}-1}{\varepsilon x},  \quad \quad g_{1}(x)=1.
\end{equation}
\end{enumerate}
Here, $\varepsilon$ is a first order dimensionless free parameter of the formalism.
In order to investigate the effect of the RG on relativistic quantum systems, in this paper we consider only the spin-zero boson solutions out of the generalized DKP oscillator solution, one should initially introduce the local reference frame, $\hat{\theta}^{a}=e^{a}_{\,\,\,\mu}(x)dx^{\mu}$, where $a=0,1,2,3$. Note that the local reference frame is associated with the line element Eq. \eqref{le2} by
\begin{equation}\label{LocalRefFram}
\begin{split}
&\hat{\theta}^{0}=\frac{dt}{g_{0}(x)}, \qquad \hat{\theta}^{1}=\frac{dr}{g_{1}(x)},\\
&\hat{\theta}^{2}=\frac{\alpha r d\varphi}{g_{1}(x)}, \qquad \hat{\theta}^{3}=\frac{dz}{g_{1}(x)}.
\end{split}
\end{equation}
Thus, via the local reference frame the components of the non-coordinate basis and their inverses, so-called tetrads and inverse tetrads, can be obtained respectively as follows:
 %At first,  we express the tetrad basis out of the cosmic string metric.
\begin{subequations}\label{tetrbas}
\begin{align}
&e^{a}_{\,\,\,\mu}(x)=\mathrm{diag}\left(\frac{1}{g_{0}(x)},\frac{1}{g_{1}(x)},\frac{\alpha r}{g_{1}(x)},\frac{1}{g_{1}(x)}\right),\label{tetrbasdown}\\
&e^{\mu}_{\,\,\,a}(x)=\mathrm{diag}\left(g_{0}(x),g_{1}(x),\frac{g_{1}(x)}{\alpha r},g_{1}(x)\right). \label{tetrbasUp}
\end{align}
\end{subequations}
It is worth noting that, tetrads must satisfy conditions  $e^{a}_{\,\,\,\mu}(x)e^{\mu}_{\,\,\,b}(x)=\delta^{a}_{\,\,\,b}$ and $e^{\mu}_{\,\,\,a}(x)e^{a}_{\,\,\,\nu}(x)=\delta^{\mu}_{\,\,\,\nu}$, while they obey the relation $\mathrm{g}_{\mu\nu}(x)=e^{a}_{\,\,\,\mu}(x)e^{b}_{\,\,\,\nu}(x)\eta_{ab}$, where the metric tensor is related to a $(1+3)$-dimensional Minkowski space-time that is denoted by the signature, $\eta_{ab}=\mathrm{diag}(-,+,+,+)$.  To continue our study, we need to convert  the partial derivative, $\partial_{\mu}$, to covariant derivative, $\nabla_{\mu}$.  For this we use the well-known definition, $\nabla_{\mu}=\partial_{\mu}-\Gamma_{\mu}(x)$, where the affine connection $\Gamma_{\mu}(x)$ is related to the DKP equation by $\Gamma_{\mu}(x)=\frac{1}{2}\omega_{\mu ab}[\beta^{a},\beta^{b}]$. Here,  $\beta^a$ matrices are the DKP matrices in Minkowski space-time.  As pointed out in Refs. \cite{ZareEJMPA2020,34,35,45,ZareGRG2020}, by using the $\beta^a$ matrices, one can exğress the DKP algebra by three irreducible representations. Among them,  a ten-dimensional representation is shown to be related to spin-one particles, while a five-dimensional representation that is related to spin-zero particles, and a one-dimensional representation that is the trivial one. Since, we consider  the spin-zero DKP field in this contribution, we have to take  the five-dimensional representation into account. While doing this, we pick out the $5\times5$ beta-matrices as
\begin{equation}\label{8}
	\beta^0=\begin{pmatrix}
		\theta&0_{2\times 3}\\
		0_{3\times 2}&0_{3\times 3}
	\end{pmatrix},\qquad
	{\vec{\beta}}=\begin{pmatrix}
		0_{2\times 2}&\vec{\tau}\\
		-\vec{\tau}^T&0_{3\times 3}
	\end{pmatrix},
\end{equation}
in which matrix transposition is denoted by $T$, and
\begin{equation}\label{9}
	\theta=\begin{pmatrix}
		0&1\\
		1&0
	\end{pmatrix},\quad \tau^1=\begin{pmatrix}
		-1&0&0\\
		0&0&0
	\end{pmatrix},\quad \tau^2=\begin{pmatrix}
		0&-1&0\\
		0&0&0
	\end{pmatrix}, \quad \tau^3=\begin{pmatrix}
		0&0&-1\\
		0&0&0
	\end{pmatrix}.
\end{equation}
Note that the DKP matrices basically resemble the Dirac matrices used in the Dirac equation \cite{47,48,49,50}, however they satisfy  a more complex algebra, namely the Kemmer’s algebra, instead:
\begin{equation}\label{DKPAlg}
	\beta^{a}\beta^{b}\beta^{c}+\beta^{c}\beta^{b}\beta^{a}=\eta^{ab}\beta^{c}+\eta^{bc}\beta^{a}, \qquad a,b,c=0,1,2,3.
\end{equation}
To find the non-null component of the affine connection, we initially should acquire the non-null components of spin connection in the lack of torsion by solving the Maurer-Cartan structure equations, $\mathrm{d}\hat{\theta}^a+\omega^{a}_{\,\,\,b}\wedge\hat{\theta}^{b}=0$, where $\omega^{a}_{\,\,\,b}=\omega^{\,a}_{\mu\,\,b}\mathrm{d}x^{\mu}$. Thereby, we get $\omega^{2}_{\varphi\,\,1}=-\omega^{1}_{\varphi\,\,2}=\alpha$. Thus, the affine connection can be written as
\begin{equation}\label{affconn}
	\Gamma_{\varphi}=
	\begin{pmatrix}
		0&0&0&0&0\\
		0&0&0&0&0\\
		0&0&0&\alpha&0\\
		0&0&-\alpha&0&0\\
		0&0&0&0&0
	\end{pmatrix}.
\end{equation}
Let us now deal with the four-vector bosonic current, $J^{\mu}=\frac{1}{2}\bar{\Psi}\beta^{\mu}\Psi$. We note that it is conserved due to  the conservation law in the following form \cite{38,38b}:
%\begin{equation}\label{consFC}
%	J^{\mu}=\frac{1}{2}\bar{\Psi}\beta^{\mu}\Psi,
%\end{equation}
%is conserved due to  the conservation law in the following form:
\begin{equation}\label{conslawFC}
	\triangledown_{\mu}J^{\mu}+\frac{i}{2}\bar{\Psi}(U-\eta^{0}U^{\dag}\eta^{0})\Psi=\frac{1}{2}\bar{\Psi}(\triangledown_{\mu}\beta^{\mu})\Psi.
\end{equation}
Here, $\bar{\Psi}$ is the adjoint spinor that is defined by  $\bar{\Psi}=\Psi^{\dag}\eta^{0}$, where $\eta^{0}=2\beta^{0}\beta^{0}-1$,  and $(\eta^{0}\beta^{\mu})^{\dag}=\eta^{0}\beta^{\mu}$.
The coefficient of $\frac{1}{2}$, which is related to $\bar{\Psi}\beta^{\mu}\Psi$, does not play a crucial role  in Eq. \eqref{conslawFC}, however, it confirms that the four-vector boson current, $J^\mu$, is compatible with that applied in the Klein-Gordon model and its non-relativistic regime. If $U$  is a Hermitian  matrice with  respect to $\eta^{0}$ and $\eta^{\mu}$, then the four-vector boson current becomes covariantly invariant and it is conserved in the following form
\begin{equation}\label{fourcurr}
	\nabla_{\mu}\beta^{\mu}=0.
\end{equation}
More than three decades ago, Moshinsky and Szczepaniak introduced an extraordinary model to represent a coupling between a harmonic oscillator and a relativistic fermion of the Dirac equation \cite{Moshinsky1989}. There, they modified the momentum operator via $\vec{p}\rightarrow \vec{p}-iM\omega r \gamma^{0} \hat{r}$, where $\gamma^{0}$ and $\hat{r}$ denotes the usual Dirac matrices and radial unit vector, respectively. It should be noted that this model is presented in an elegant way that keeps the linearity of the Dirac equation in both momenta and spatial coordinates. This model retrieves the Schrödinger equation of a harmonic oscillator out of the Dirac equation in the nonrelativistic regime. Based on such a valuable work, we were motivated to examine the generalized DKP equation by adding a non-minimal coupling as it is done in the Dirac oscillator.
Hereafter, we will call the  DKP equation under the presence of the relevant non-minimal coupling as the generalized DKP oscillator (gDKPo). Due to this non-minimal coupling, the covariant derivative changes to $\nabla_{\mu}\rightarrow \partial_{\mu}-\Gamma_{\mu}(x)+  M \omega \eta^{0} r \delta ^{r}_{\mu}$ \cite{ZareEJMPA2020,37,46,47a}, where the oscillator frequency is denoted by $\omega$ and the mass of the spin-0 boson is expressed with $M$. Thus, the gDKPo in the considered cosmic background can be demonstrated as
\begin{equation}\label{DKPEQ}
\left[i\beta^{\mu}\left(\partial_{\mu}+ M \omega \eta^{0} r-\Gamma_{\mu}(x)\right)-M\right]\Psi\left(t,\vec{r}\right)=0.
\end{equation}
Here, DKP field is demonstrated by $\Psi\left(t,\vec{r}\right)$ and the generalized DKP matrices in this background are indicated by $\beta^{\mu}$. Then, we express the beta matrices that correspond to their flat space-time counterparts
\begin{equation}\label{defbeta}
\beta^{\mu}=e^{\mu}_{\,\,\,a}\,\beta^{a}.
\end{equation}
Next, we employ the specific tetrads bases which are represented by Eq. \eqref{tetrbasUp} in Eq. \eqref{defbeta}, and we find the generalized DKP matrices in terms of the usual DKP matrices as follows:
\begin{equation}\label{CurvSpacBeta}
\begin{split}
&\beta^{t}=e^{t}_{a}\beta^{a}=g_{0}\beta^{0}, \qquad \qquad
\beta^{r}=e^{r}_{a}\beta^{a}=g_{1}\beta^{1},\\
&\beta^{\varphi}=e^{2}_{a}\beta^{a}=\frac{g_{1}\beta^{2}}{r\alpha},\qquad \qquad
\beta^{z}=e^{z}_{a}\beta^{a}=g_{1}\beta^{3}.
\end{split}
\end{equation}
In this manuscript, we consider time-independent interactions, therefore we propose that the wave function to be in the form of  $\Psi\left(t,\vec{r}\right)\propto  e^{-iEt}e^{im\varphi+ik_{z}z}\Phi(r)$. Here, we use $E$ to denote the energy of the bosonic particle, $m$ to indicate the magnetic quantum number, and $k_{z}$ to express the wave number. $\Phi(r)$ is the five-component DKP spinor in such a way that its transpose is given by \linebreak $\Phi^{T}=(\Phi_{1}, \Phi_{2}, \Phi_{3}, \Phi_{4}, \Phi_{5})$. Therefore, to solve Eq. \eqref{DKPEQ}, we substitute the proposed wave function $\Psi\left(t,\vec{r}\right)$ in  Eq. \eqref{DKPEQ}
and arrive at
\begin{subequations}\label{DKPEq}
\begin{align}
-&r\alpha {M}\Phi_{1}(r)+Eg_{0}(x)r\alpha \Phi_{2}(r)-i\alpha g_{1}(x)\left(1+r\partial_{r}-M\omega r^{2} \right)\Phi_{3}(r)\nonumber\\	
+&g_{1}(x)m	\Phi_{4}(r)+g_{1}(x)k_{z}r \alpha \Phi_{5}(r)=0.\label{DKPEq1}\\
%-&r\alpha {M}\Phi_{1}(r)+Eg_{0}(x)r\alpha \Phi_{2}(r)+g_{1}(x)m\Phi_{4}(r)\nonumber\\
%+&g_{1}(x)\alpha( {i}(-1+r^{2}{M}\omega)\Phi_{3}(r)+r(k_{z}\Phi_{5}(r)-{i}\Phi'_{3}(r)))=0, \label{DKPEq1}\\
&Eg_{0}(x)\Phi_{1}(r)-{M}\Phi_{2}(r)=0,\label{DKPEq2}\\
&\left(g_{1}(x)r{M}\omega+g_{1}(x)\partial_{r}\right)\Phi_{1}(r)+{iM}\Phi_{3}(r)=0\label{DKPEq3},\\
-&g_{1}(x)m\Phi_{1}(r)-r\alpha {M}\Phi_{4}(r)=0,\label{DKPEq4}\\
&g_{1}(x)k_{z}\Phi_{1}(r)+{M} \Phi_{5}(r)=0.\label{DKPEq5}
\end{align}
\end{subequations}
By solving the coupled equation system of  \eqref{DKPEq} in  terms of   $\Phi_{1}$, we obtain
\begin{subequations}\label{spnDKP}
\begin{eqnarray}
\Phi_{2}(r)&=&\frac{Eg_{0}(x)}{M}\Phi_{1}(r)\label{spnDKP2},\\
\Phi_{3}(r)&=&\frac{ig_{1}(x)}{M}\left(r M\omega\Phi_{1}(r)+\partial_{r}\Phi_{1}(r) \right)\label{spnDKP3},\\
\Phi_{4}(r)&=&-\frac{g_{1}(x)m}{\alpha M}\frac{\Phi_{1}(r)}{r}\label{spnDKP4},\\
\Phi_{5}(r)&=&-\frac{g_{1}(x)k_{z}}{ M}\Phi_{1}(r)\label{spnDKP5}.
\end{eqnarray}
\end{subequations}
In the following, by substituting Eq. \eqref{spnDKP} in Eq. \eqref{DKPEq1}, a second-orders radial differential equation
can be written in terms of the first component of the gDKPo,  that is, $\Phi_{1}(r)$ containing the RG functions proposed in Eqs. \eqref{RFm1}, \eqref{RFm2} and \eqref{RFm3} in the form
\begin{equation}\label{DKPwaveEq1}
\begin{split}
\frac{d^{2}\Phi_{1}(r)}{dr^2}+\frac{1}{r}\frac{d\Phi_{1}(r)}{dr}+\left[\frac{g_{0}(x)}{g_{1}(x)}E^{2}-\frac{M^{2}}{g_{1}(x)}+2M\omega-\frac{m^{2}}{\alpha^{2}r^{2}}-M^{2}\omega^{2}r^{2}-k_{z}^{2}\right]\Phi_{1}(r)=0.
\end{split}
\end{equation}
In search of solutions of the gDKPo, we assume $k_{z}=0$ that it leads to the elimination of the fifth component of the equation, that is, $\Phi_{5}(r)=0$. In this regard, to solve Eq. \eqref{DKPwaveEq1}, we need to utilize each pair of these RG functions Eqs. \eqref{RFm1}, \eqref{RFm2} and \eqref{RFm3} and explore the associated solutions in the following cases.
\begin{enumerate}
\item
{\bf \large The first case: }\\
Initially, we examine the first chosen form of the RG functions. Therefore, we employ Eq. \eqref{RFm1}
in Eq. \eqref{DKPwaveEq1} and arrive at an equation of motion for $\Phi_1(r)$ as
\begin{equation}\label{secorddeq1}
\left[\frac{d^2}{dr^{2}}+\frac{1}{r}\frac{d}{dr}+\left(E^{2}_{{nm}}-(1-x \varepsilon)^{2}{{M}^{2}}
+2{{M}} \omega - \frac{m^{2}}{r^{2}\alpha^{2}}-r^{2}{{M}^{2}{\omega^{2}}}\right)\right]\Phi_{1_{{nm}}}(r)=0.
\end{equation}
Next, we define a new variable, namely $M\omega\, r^{2}=\rho$, then, Eq. \eqref{secorddeq1}  reduces to
\begin{equation}\label{secorddeq2}
\left[\rho \frac{d^{2}}{d\rho^{2}}+\frac{d}{d\rho}-\left(\frac{j^{2}}{4\rho}+\frac{\rho}{4}-\frac{{\kappa}^{2}}{4M \omega}\right)\right]\Phi_{1_{nm}}(\rho)=0,
\end{equation}
where
\begin{subequations}\label{parajk}
\begin{align}
&j^{2}=\frac{m^{2}}{\alpha^{2}},\\
&\kappa^{2}=E_{{nm}}^{2}-(1-x\varepsilon)^{2}M^{2}+2M\omega.  \end{align}
\end{subequations}
In the way of solving
Eq. \eqref{secorddeq2}, the following wave function can be suggested
\begin{equation}\label{WFEq1}
{\Phi_{1_{nm}}}(\rho) =\rho^{\frac{|j|}{2}}e^{-\frac{\rho}{2}}F_{{nm}}(\rho).
\end{equation}
According to Eq. \eqref{WFEq1}, we rearrange Eq. \eqref{secorddeq2} in the form
\begin{equation}\label{secorddeq3}
\rho \frac{d^2F_{nm}(\rho)}{d\rho^2}+\left(1+|j|-\rho\right)\frac{dF_{nm}(\rho)}{d\rho}-\left(\frac{1+|j|}{2}-\frac{\kappa^{2}}{4M\omega}\right)F_{nm}(\rho)=0.
\end{equation}
{The general solution of Eq. \eqref{secorddeq3} is the confluent hypergeometric function \cite{51,52} which can be written in the form of}
\begin{equation}\label{HY1F1}
F_{nm}(\rho)={}_{1}F_{1}\left(\frac{1+|j|}{2}-\frac{\kappa^{2}}{4M\omega}, 1+j,\rho\right).
\end{equation}
{Thus, the radial wave function of the second-order differential Eq. \eqref{secorddeq2} can be arranged as}
\begin{equation}\label{WF1}
\Phi_{1_{nm}(\rho) }=\rho^{\frac{|j|}{2}}e^{-\frac{\rho}{2}}\,\,{}_{1}F_{1}\left(\frac{1+|j|}{2}-\frac{\kappa^{2}}{4M\omega}, 1+j,\rho\right).\\
\end{equation}
Accordingly, we obtain the DKP spinor $\Phi(r)$ as
\begin{equation}\label{DKPSpinor1}
\Phi_{nm}(r)=N_{nm}
\begin{pmatrix}
1\\
\frac{E}{M\left(1-\varepsilon x\right)}\\
\frac{i\left(r M\omega+\partial_{r}\right)}{M\left(1-\varepsilon x\right)}\\
-\frac{m}{r\alpha M\left(1-\varepsilon x\right)}\\
0
\end{pmatrix}\left(M\omega\, r^{2}\right)^{\frac{|j|}{2}}e^{-\frac{M\omega\, r^{2}}{2}}\,\,{}_{1}F_{1}\left(\frac{1+|j|}{2}-\frac{\kappa^{2}}{4M\omega}, 1+j,M\omega\, r^{2}\right).
\end{equation}
If $F_{nm}(\rho)$ function is a polynomial of degree $n$, we can present the solutions of the bound states, because of the divergent behavior of this function in large values of its argument. The confluent hypergeometric function which is indicated by ${}_{1}F_{1}(\alpha; \gamma; z)$ is given by the following definition: ${}_{1}F_{1}(\alpha; \gamma; z)=\sum_{n=0}^{\infty}\frac{(\alpha)_{n}}{n!(\gamma)_{n}}z^{n}$, with $\alpha=\frac{1+|j|}{2}-\frac{\kappa^{2}}{4M\omega}$, $\gamma=1+j$ and $z=M\omega r^{2}$, where $\gamma\ne0,-1,-2,\dots$ and $(\alpha)_{n}=\frac{\Gamma(\alpha+n)}{\Gamma(\alpha)}$ and also around the origin $z\ll 1$. Therefore, ${}_{1}F_{1}(\alpha; \gamma; z)\simeq 1+\frac{\alpha}{\gamma}z+O(z^{2})$. By the way, the asymptotic behavior for large $|z|$ is: ${}_{1}F_{1}(\alpha; \gamma; z)\simeq\frac{\Gamma(\gamma)}{\Gamma(\alpha)}e^{z}z^{\alpha-\gamma}\left[1+O(|z|^{-1})\right]$ (here we assume that $z\rightarrow \infty$).

Here,  the radial wave function Eq. \eqref{WF1} shows an admissible behavior at infinity. This condition is written by
\begin{equation}\label{Cond1}
\frac{1+|j|}{2}-\frac{\kappa^{2}}{4M\omega}=-n, \quad \quad  \quad \quad n=0,1,2,3,....
\end{equation}
It is worth mentioning that from Eq. \eqref{Cond1} one can produce the quantization condition on the energy spectrum of the particle. In this case the explicit form of energy is given by
\begin{eqnarray}\label{EnergM1}
\frac{E_{nm}}{E_P}&=&\frac{-\varepsilon\big(\frac {M}{E_P}\big)^2\mp\sqrt{\big(\frac {M}{E_P}\big)^2+\Big(1-\varepsilon^2\big(\frac {M}{E_P}\big)^2\Big)\Big(\frac{M}{E_P}\Big)\Big(\frac{2|m|}{\alpha}+4n\Big)\Big(\frac{\omega}{E_P}\Big)}}{\Big(1-\varepsilon^2\big(\frac {M}{E_P}\big)^2\Big)}
\end{eqnarray}
Then, we present the effect of the angular parameter on the energy spectrum. We take an arbitrary energy eigenvalue function, $E_{11}$, and depict its ratio to the Planck energy versus the ratio of the oscillator frequency to the Planck energy according to three different values of deficit angles in Fig. \ref{Fig1}. We observe that for higher deficit angles the forbidden energy width becomes narrow. On the other hand for a fixed value of deficit angle, the increase of the oscillator frequency widens the gap of the energy eigenvalues.
\begin{figure}[H]
	\centering
	\subfigure[For $\frac{M}{E_{P}}=0.1$]{%
		\label{fig41}%
		\includegraphics[height=7cm,width=8cm]{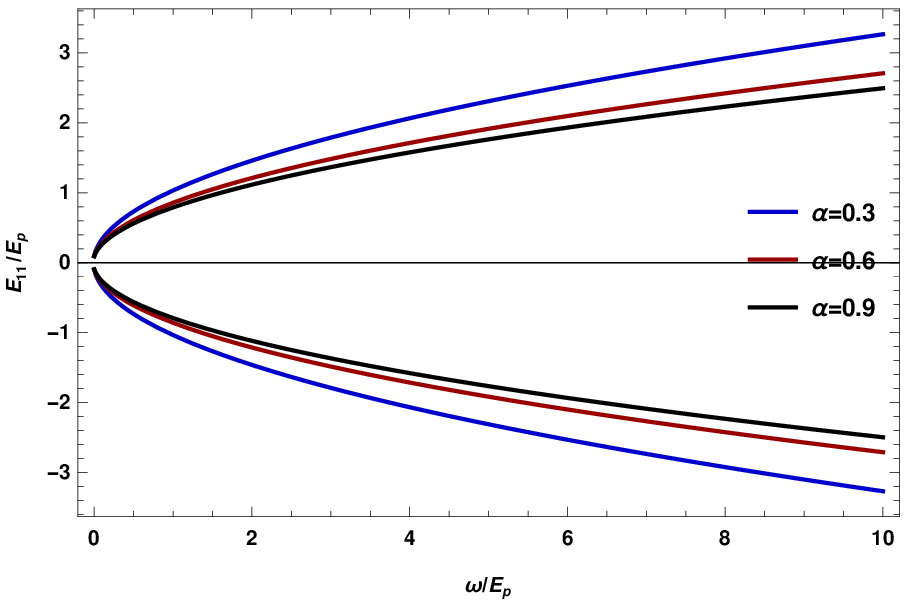}}%
	\qquad
	\subfigure[For $\frac{M}{E_{P}}=0.8$]{%
		\label{fig42}%
		\includegraphics[height=7cm,width=8cm]{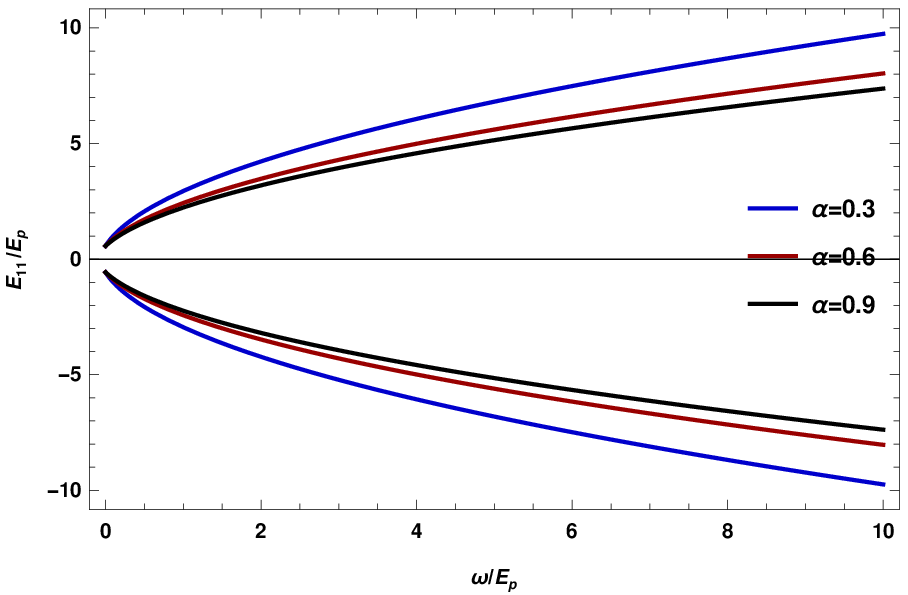}}%
	\caption{The reduced energy eigenvalue function $(\frac{E_{nm}}{E_P})$ versus the reduced oscillator frequency $(\frac{\omega}{E_P})$ with $\varepsilon=0.5$, $m=1$ and $n=1$ for different $\alpha$ values.}
	\label{Fig1}
\end{figure}
%\begin{figure}[H]
%	\begin{center}
%		\includegraphics[scale=1.0]{EOmeg1new.eps}
%		\caption{Reduced energy eigenvalue function versus reduced oscillator frequency, $\frac{M}{E_{P}}=0.8;\varepsilon=0.5; m=1; n=1$. \label{Fig1}}
%	\end{center}
%\end{figure}
Next, we derive the probability density function in this scenario. We use  Eq. \eqref{DKPSpinor1} in the  four-vector bosonic current. We find
\begin{equation}\label{fourcurrent1}
J^{t}=\frac{1}{2}\bar{\Psi}\beta^{t}\Psi=\frac{E_{nm}}{\left(1-\varepsilon x\right)M}\left|\Phi_{1_{nm}}(r)\right|^2.
\end{equation}
Then, for three different values of deficit angles we plot  $J^{t}/E_{p}$
versus $r/E_{p}$ in Fig. \ref{Fig2} with a fixed energy eigenvalue.
\begin{figure}[H]
	\centering
	\subfigure[For $\frac{M}{E_{P}}=0.2$]{%
		\label{fig21}%
		\includegraphics[height=7cm,width=8cm]{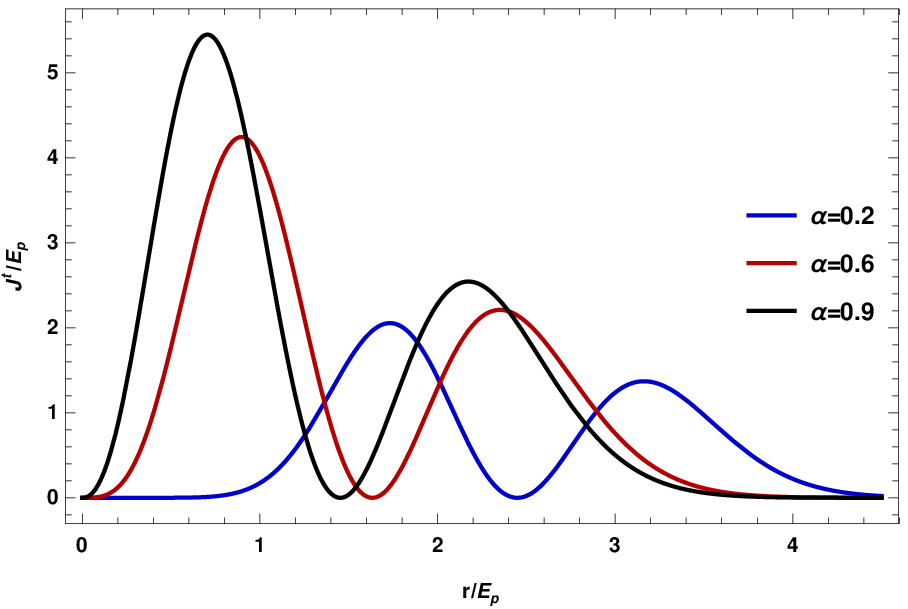}}%
	\qquad
	\subfigure[For $\frac{M}{E_{P}}=0.8$]{%
		\label{fig22}%
		\includegraphics[height=7cm,width=8cm]{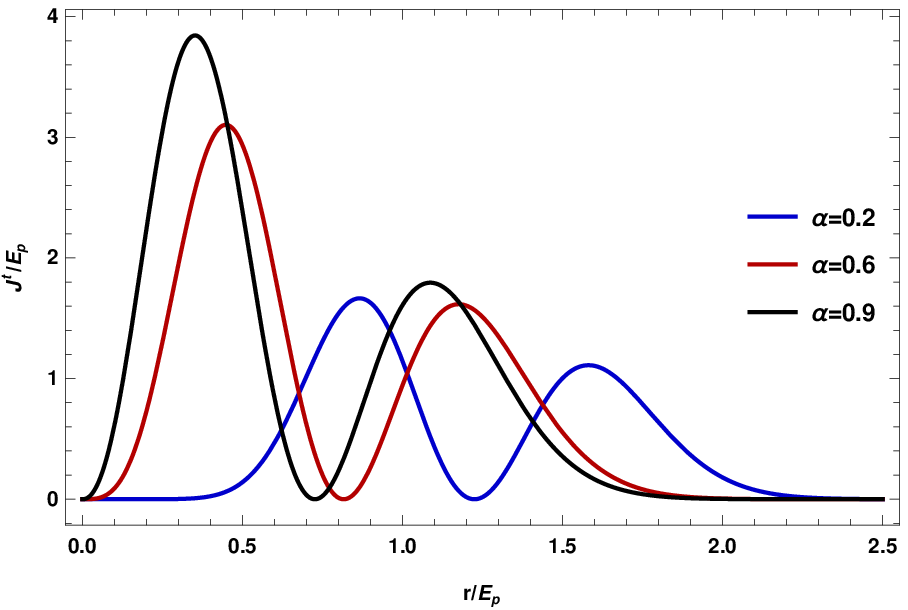}}%
	\caption{The reduced probability density function $(\frac{J^{t}}{E_P})$ versus the reduced spatial distance $(\frac{r}{E_P})$ with  $\varepsilon=1$, $m=1$, $n=1$ and $\omega=5$ for different $\alpha$ values.}
	\label{Fig2}
\end{figure}
%\begin{figure}[H]
%\begin{center}
%\includegraphics[scale=1.0]{JtRF1.eps}
%\caption{Reduced probability density  function  versus the reduced spatial distance  with $\frac{M}{E_{P}}=0.8$, $\varepsilon=1$, $m=1$, $n=1$ and $\omega=5$ for different $\alpha$.\label{Fig2}}
%\end{center}
%\end{figure}

\item

{\bf \large The second case:} \\
{We continue our research by examining the second chosen form of the RG functions. Similar to the first case, we substitute Eq. \eqref{RFm2} in  Eq. \eqref{DKPwaveEq1}  to derive an equation of motion in terms of the first component of the DKP spinor. We arrive at
\begin{eqnarray}\label{secorddeq4}
\left[\frac{d^2}{dr^{2}}+\frac{1}{r}\frac{d}{dr}+\left(\frac{E^{2}_{nm}-M^{2}}{1-\varepsilon\, x^{2}}+2M\omega - \frac{m^{2}}{r^{2}\alpha^{2}}-r^{2}M^{2}\omega^{2}\right)\right]\Phi_{1_{nm}}(r)=0.
\end{eqnarray}}
{We have to remark that this equation is in the same form with Eq. \eqref{secorddeq1}. Therefore, its general solution has to be in the same form. We avoid to repeat the same steps and and express the only difference in the solution. The confluent hypergeometric function, given in Eq. \eqref{HY1F1},  has a different parameter, namely $\kappa'^{2}$ instead of $\kappa^{2}$.}
\begin{equation}\label{kprim}
\kappa'^{2}=\frac{E^{2}_{nm}-M^{2}}{1-\varepsilon\, x^{2}}+2M\omega.
\end{equation}
After applying the quantization condition, we obtain the following  energy eigenvalue expression:
{\begin{eqnarray}\label{EnergM2}
\frac{E_{nm}}{E_P}&=&\mp\sqrt{\frac{\big(\frac{M}{E_P}\big)^2+2\big(\frac{M}{E_P}\big)\big(\frac{\omega}{E_P}\big)\big(\frac{|m|}{\alpha}+2n\big)}{1+2\varepsilon \big(\frac{M}{E_P}\big)\big(\frac{\omega}{E_P}\big)\big(\frac{|m|}{\alpha}+2n\big)}}
\end{eqnarray}}
The energy eigenvalue function is symmetric, unlike the first case's energy eigenfunction. Moreover, we observe that at the higher oscillator frequencies energy eigenvalues converge to a constant, namely inverse square root of $\varepsilon$. We would like to emphasize that in the first case we did not see any convergence.  Then, we present the plot of the reduced energy eigenvalues $(\frac{E_{nm}}{E_P})$ versus the reduced oscillator frequency $(\frac{\omega}{E_P})$ in Fig. \ref{Fig3}. Alike in the first case, we use three different angular parameters. We observe that in this case, the energy functions do not alter as much as they do in the first case.
\begin{figure}[H]
\begin{center}
\includegraphics[scale=1.1]{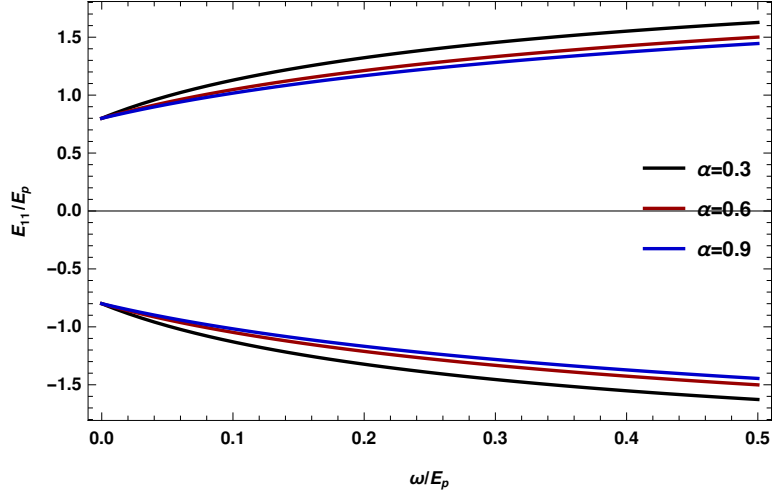}
\caption{The reduced energy eigenvalue function $(\frac{E_{nm}}{E_P})$ versus the reduced oscillator frequency $(\frac{\omega}{E_P})$ with $\frac{M}{E_{P}}=0.8$, $\varepsilon=0.2$, $m=1$ and $n=1$.\label{Fig3}}
\end{center}
\end{figure}	
{Before we study the third case, we intend to examine the probability density function. Therefore, we use the first component of the DKP spinor
\begin{equation}\label{WF2}
\Phi_{1_{nm}(\rho) }=\rho^{\frac{|j|}{2}}e^{-\frac{\rho}{2}}\,\,{}_{1}F_{1}\left(\frac{1+|j|}{2}-\frac{{\kappa'}^{2}}{4M\omega}, 1+j,\rho\right),\\
\end{equation}
and obtain the probability density function as}
\begin{equation}\label{fourcurrent2}
J^{t}=\frac{E_{nm}}{M}\left|\Phi_{1_{nm}}(r)\right|^2.
\end{equation}
{Then, we use Eq. \eqref{fourcurrent2} to depict the reduced probability density $(\frac{J^{t}}{E_P})$ versus the reduced spatial coordinate $(\frac{r}{E_P})$ in Fig. \eqref{Fig4}. Here, unlike the first case, we keep the deficit angle as a constant and employ three different values of oscillator frequency. }
\begin{figure}[H]
\begin{center}
\includegraphics[scale=1.1]{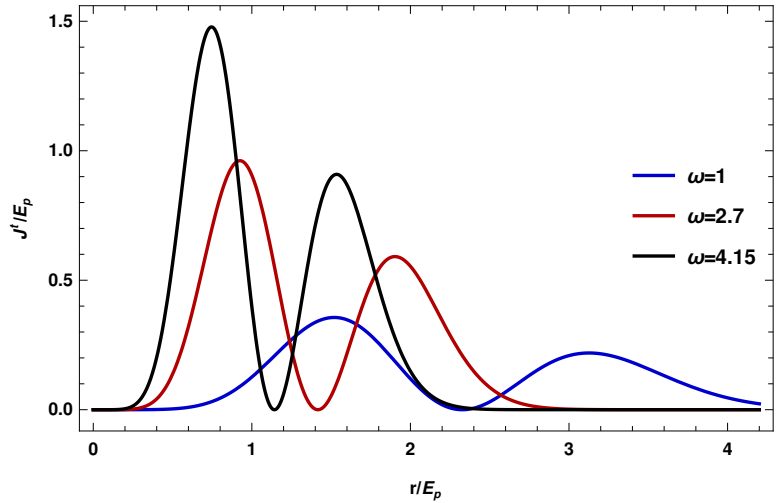}
\caption{The reduced probability density versus the reduced spatial coordinate  with $\frac{M}{E_{P}}=0.8$, $\varepsilon=1$, $m=1$, $n=1$ and $\alpha=0.3$ for different $\omega$.\label{Fig4}}
\end{center}
\end{figure}

\item	
{\bf \large The third case:} \\
{Finally, we examine the third scenario that is presented in Eq. \eqref{RFm3}. In this case, one of the RG functions is assumed to be proportional to an exponential function. Analogous to the previous two cases, we use the RG functions in Eq. \eqref{DKPwaveEq1} to obtain an equation of motion in terms of the first component of the DKP spinor component. So, we find
\begin{eqnarray}\label{secorddeq5}
\left[\frac{d^2}{dr^{2}}+\frac{1}{r}\frac{d}{dr}+\left(\frac{(1-e^{\varepsilon x})^{2}E^{2}}{(\varepsilon x)^{2}}-M^{2}+2M\omega-\frac{m^{2}}{r^{2}\alpha^{2}}-r^{2}M^{2}\omega^{2}\right)\right]\Phi_{1_{nm}}(r)=0.
\end{eqnarray}}
The solution of Eq. \eqref{secorddeq5} is similar to the solutions in the previous section and it is written in terms of the confluent hypergeometric function. The only difference is the use of a  new parameter, $\tilde{\kappa}$, instead of  $\kappa$
{\begin{equation}\label{kzeg}
\tilde{\kappa}^{\,2}=\frac{(1-e^{\varepsilon x})^{2}E^{2}}{(\varepsilon x)^{2}}-M^{2}+2M\omega.
\end{equation}
 has to be used. We repeat the algebraic steps and  we find the energy eigenvalue function in the form of
\begin{eqnarray}\label{EnergyEq3}
\frac{E_{nm}}{E_P}&=& \frac{1}{\varepsilon}\ln{\left[1\mp\varepsilon\sqrt{\left(\frac{M}{E_P}\right)^{2}+ 2\left(\frac{M}{E_P}\right)\left(\frac{\omega}{E_P}\right)\left(\frac{|m|}{\alpha}+2n\right)}\right]}
\end{eqnarray}}
{Then, we demonstrate the energy eigenfunction solutions. At first, in Fig. \ref{Fig5}, we present the solution with the positive sign.  We observe that the energy function increases with a decreasing increase. Moreover, for higher angular parameter this increase become smaller. }
 \begin{figure}[H]
\begin{center}
\includegraphics[scale=1.1]{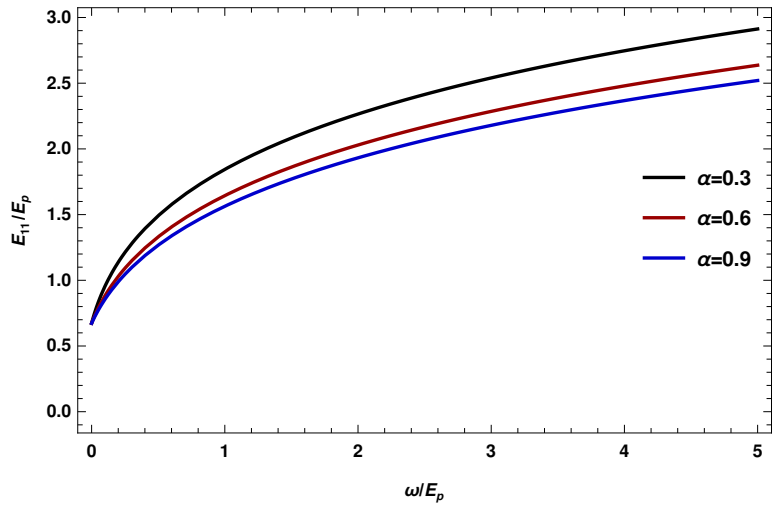}
\caption{The reduced energy eigenvalue function $(\frac{E_{nm}}{E_P})$ versus the reduced oscillator frequency $(\frac{\omega}{E_P})$ with $\frac{M}{E_{P}}=0.8$, $\varepsilon=0.5$, $m=1$ and  $n=1$.\label{Fig5}}
\end{center}
\end{figure}
{ Next, in Fig. \ref{Fig6}, we illustrate the second solution, namely the solution with the negative sign. We observe that the energy eigenvalues diverge at a certain cut-off frequency. We note that this frequency value depends on the value of the deficit angle. More precisely, we obtain a higher cut-off frequency value with a higher deficit parameter. }
\begin{figure}[H]
\begin{center}
\includegraphics[scale=1.1]{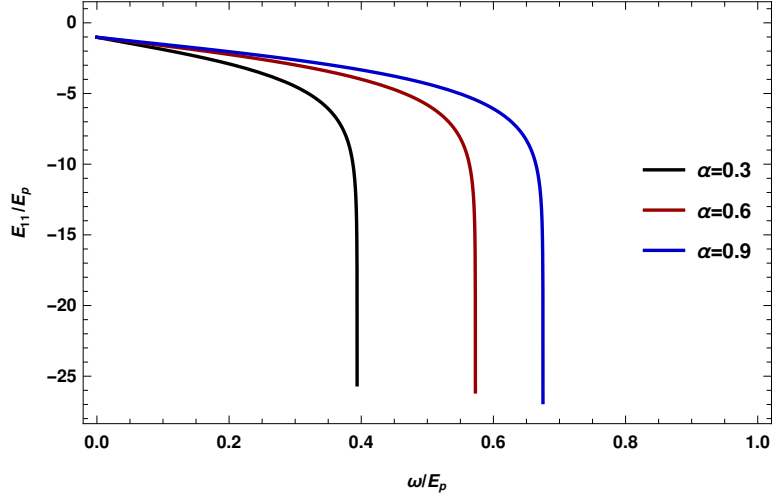}
\caption{The reduced energy eigenvalue function $(\frac{E_{nm}}{E_P})$ versus the reduced oscillator frequency $(\frac{\omega}{E_P})$ with $\frac{M}{E_{P}}=0.8$, $\varepsilon=0.5$, $m=1$ and $n=1$.\label{Fig6}}
\end{center}
\end{figure}
The presence of a critical value of cut-off frequency gives the hint of a symmetry breaking in this kind of rainbow gravity scenarios. Before we finish the examination, we would like to express the first component of the DKP spinor.
\begin{equation}\label{WF3}
\Phi_{1_{nm}(\rho) }=\rho^{\frac{|j|}{2}}e^{-\frac{\rho}{2}}\,\,{}_{1}F_{1}\left(\frac{1+|j|}{2}-\frac{{\tilde{\kappa}}^{2}}{4M\omega}, 1+j,\rho\right).\\
\end{equation}
Then, the relevant probability density function becomes
\begin{equation}\label{fourcurrent3}
J^{t}=\frac{\left(e^{\varepsilon x}-1\right)E_{nm}}{\varepsilon x M}\left|\Phi_{1_{nm}}(r)\right|^2.
\end{equation}
Eq. \eqref{fourcurrent3} shows that the probability density increases in terms of $x=\frac{E}{E_{p}}$ more than $\left|\Phi_{1_{nm}}(r)\right|^2$.

\end{enumerate}
%%%%%%%%%%%%%%%%%%%%%%%%  CONCLUSION

\section{Conclusion\label{Conclusion}}

{The overall objective of this paper is to examine dynamics of a spin-0 boson particle that is assumed to be under the influence of a  DKP oscillator field  in  a cosmic string space-time within the context of three different rainbow  gravity scenarios.  We obtained energy eigenvalue functions and their corresponding DKP spinors analytically for  each pairs of rainbow gravity functions which are used to modify the dispersion relation. Although the first selected function pair is equivalent to each other and therefore symmetrical, the derived energy eigenvalue function did not present a symmetrical property with respect to zero energy. In this case, we observed that the deficit parameter tunes the forbidden gap width. In the second case, the rainbow functions were not symmetric to each other, however, the derived energy eigenfunction showed a symmetrical property with respect to null energy. We observed that the increase of the oscillator frequency do not affect the energy eigenvalue function after it reaches to a certain value that is proportional to the inverse square root of the  rainbow function free parameter. In the third case, we employ an exponential rainbow function. We found that one of the root of the energy eigenvalue function become unphysical after a critical cutoff oscillator frequency. We concluded that this can be seen as a symmetry breaking in that scenario. We supported our findings with the graphs of energy eigenvalues and probability densities. }

\section*{Acknowledgment}
The authors thank the referee for a thorough reading of our manuscript and for constructive suggestion. The authors would like to thank Soroush Zare for his help. This work is supported by the Internal Project, [2021/2212], of Excellent Research of the Faculty of Science of University Hradec Kr\'alov\'e.

\end{document}